\documentclass[journal]{IEEEtran}

\usepackage{amsmath}
\usepackage{graphicx}
\usepackage{mathtools}
\usepackage{amsfonts}
\usepackage{pifont}
\usepackage{amssymb}
\usepackage{epstopdf}
\usepackage{color}
\usepackage[utf8]{inputenc}
\usepackage{setspace}
\usepackage{ragged2e}
\usepackage{epsfig}
\usepackage{tabu}

\makeatletter
\newcommand{\vast}{\bBigg@{1.2}}
\newcommand{\Vast}{\bBigg@{2.3}}
\newcommand{\vastl}{\bBigg@{4}}
\newcommand{\Vastl}{\bBigg@{5}}
\newcolumntype{M}[1]{>{\centering\arraybackslash}m{#1}}
\newcolumntype{N}{@{}m{0pt}@{}}

\newtheorem{proposition}{Proposition}

\makeatother
\ifCLASSINFOpdf
 
\else
  
\fi

\begin{document}

\title{Selection Combining Scheme over Non-identically Distributed Fisher-Snedecor $\mathcal{F}$ Fading Channels}

\author{Hussien Al-Hmood,~\IEEEmembership{Member,~IEEE,} 
         and H. S. Al-Raweshidy,~\IEEEmembership{Senior Member,~IEEE}
\thanks{Manuscript received May 6, 2019; xxxxx xxxxx xxxxx xxxxx xxxxx xxxxx xxxxx xxxxx xxxxx xxxxx xxxxx xxxxx xxxxx xxxxx xxxxx xxxxx xxxxx xxxxx xxxxx xxxxx xxxxx xxxxx xxxxx xxxxx xxxxx xxxxx xxxxx  xxxxx xxxxx xxxxx  xxxxx xxxxx xxxxx  xxxxx xxxxx xxxxx xxxxx xxxxx.}
\thanks{Hussien Al-Hmood is with the Department of Electronic and Computer Engineering, College of Engineering, Design and Physical Sciences, Brunel University London, UB8 3PH, U.K., e-mails: hussien.al-hmood@brunel.ac.uk, h.a.al-hmood@ieee.org.}
\thanks{H. S. Al-Raweshidy is with the Department of Electronic and Computer Engineering, College of Engineering, Design and Physical Sciences, Brunel University London, UB8 3PH, U.K., e-mail: hamed.al-raweshidy@brunel.ac.uk.}}

\markboth{IEEE WIRELESS COMMUNICATIONS LETTERS,~Vol.~00, ~NO.~00, April 2019}%
{Author 1 \MakeLowercase{\textit{et al.}}: Bare Demo of IEEEtran.cls for Journals}

\maketitle

\begin{abstract}
In this paper, the performance of the selection combining (SC) scheme over Fisher-Snedecor $\mathcal{F}$ fading channels with independent and non-identically distributed (i.n.i.d.) branches is analysed. The probability density function (PDF) and the moment generating function (MGF) of the maximum i.n.i.d. Fisher-Snedecor $\mathcal{F}$ variates are derived first in terms of the multivariate Fox's $H$-function that has been efficiently implemented in the technical literature by various software codes. Based on this, the average bit error probability (ABEP) and the average channel capacity (ACC) of SC diversity with i.n.i.d. receivers are investigated. Moreover, we analyse the performance of the energy detection that are widely employed to perform the spectrum sensing in cognitive radio networks via deriving the average detection probability (ADP) and the average area under the receiver operating characteristics curve (AUC). To validate our analysis, the numerical results are affirmed by the Monte Carlo simulations.
\end{abstract}
\IEEEpeerreviewmaketitle
\section{Introduction}
\IEEEPARstart{T}{o} mitigate the impacts of the multipath fading and shadowing on the performance of wireless communications systems, diversity reception techniques have been used in the open technical literature. Selection combining (SC) approach has been considered as an efficient diversity scheme to improve the signal-to-noise-ratio (SNR) at the receiver side. This is because it's a non-coherent combining technique where the branch with a high SNR is selected among many branches [1]. The statistical properties, namely, the probability density function (PDF), the cumulative distribution function (CDF), and the moment generating function (MGF), of the maximum of random variables (RVs) of the fading channels are widely employed to study the SC diversity [2]-[5]. In this context, the SC receivers over independent and non-identically distributed (i.n.i.d.) generalized $K_G$ fading channels was investigated in [2]. The authors in [3] studied the average bit error probability (ABEP) of SC technique with i.n.i.d. branches over $\kappa-\mu$ shadowed fading channels. In [4], the PDF, the CDF, and the MGF of the maximum of $\eta-\mu$/gamma RVs were derived and used in the analysis of average channel capacity (ACC) of wireless communications systems. Based on the results of [4], the behaviour of energy detection (ED) that is one of the most utilised spectrum sensing methods was analysed in [5] by providing unified expressions for the average detection probability (ADP) and the average area under the receiver operating characteristics (ROC) curve (AUC).  
\par More recently, the Fisher-Snedecor $\mathcal{F}$ fading channel has been proposed as a composite of Nakagami-$m$/inverse Nakagami-$m$ distributions to model device-to-device (D2D) fading channels at 5.8 GHz in both indoor and outdoor environments [6]. In contrast to the generalised-$K$ fading channel, the statistics of the Fisher-Snedecor $\mathcal{F}$ fading channel are expressed in simple analytic functions. Furthermore, it includes Nakagami-$m$, Rayleigh, and one-sided Gaussian as special cases. In addition, the Fisher-Snedecor $\mathcal{F}$ fading channel can be utilised for both line-of-sight (LoS) and non-LoS (NLoS) communications scenarios with better fitting to the empirical measurements than the generalised-$K$ ($K_G$) fading model. The authors in [7] derived the basic statistics of the sum of i.n.i.d. Fisher-Snedecor $\mathcal{F}$ RVs with applications to maximal ratio combining (MRC) receivers. The ADP and the average AUC of ED with square law selection (SLS) branches over arbitrarily distributed Fisher-Snedecor $\mathcal{F}$ fading channels were given in [8]. The product of multiple Fisher-Snedecor $\mathcal{F}$ RVs, namely, cascaded fading model, was addressed in [9]. 
\par To the best authors' knowledge, the statistical characteristics of the maximum of i.n.i.d. Fisher-Snedecor $\mathcal{F}$ variates have not been yet reported in the open literature. Motivated by this and based on the above observations, this paper derives exact analytic closed-form mathematically tractable of the PDF and the MGF of the maximum of i.n.i.d. Fisher-Snedecor $\mathcal{F}$ RVs. To this end, the performance of SC scheme is analysed by deriving the ABEP, the ACC, the ADP and the average AUC of ED in terms of the multivariate Fox's $H$-function. 

\section{The PDF and MGF of the Maximum I.N.I.D. Fisher-Snedecor $\mathcal{F}$ Variates}
The CDF of the received instantaneous SNR, $\gamma$, at $i$th branch of a SC receiver over Fisher-Snedecor $\mathcal{F}$ fading channel is expressed as [6, eq. (11)] 
\label{eqn_1}
\begin{align}
F_{\gamma_i}(\gamma)&=\frac{\Xi_i^{m_i} \gamma^{m_i}}{m_i B(m_i,m_{s_i})} {_2F_1(m_i+m_{s_i},m_i;1+m_i;-\Xi_i \gamma)}
\end{align}
where $\Xi_i=\frac{m_i}{m_{s_i} \bar{\gamma}_i}$, for $i=1,\cdots,L$, $m_i$, $m_{s_i}$, $L$, and $\bar{\gamma}_i$ stand for the multipath index, the shadowing parameter, the number of diversity branches, and  the average SNR, respectively, $B(.,.)$ is the beta function [10, eq. (8.380.1)] and  $_2F_1(.,.;.;.)$ is the Gauss hypergeometric function [10, eq. (9.14.1)].  
\par Recalling the identity [11, eq. (1.132)] and performing some mathematical simplifications with the aid of [10, eq. (8.384.1)] and [10, eq. (8.331.1)], (1) can be equivalently rewritten as 
\label{eqn_2}
\begin{align}
F_{\gamma_i}(\gamma)=&\frac{\Xi_i^{m_i} \gamma^{m_i} }{\Gamma(m_i) \Gamma(m_{s_i})} \nonumber\\
&\times H^{1,2}_{2,2} 
\bigg[ \Xi_i \gamma \bigg\vert
\begin{matrix}
  (1-m_i-m_{s_i},1), (1-m_i,1)\\
  (0,1),(-m_i,1)\\
\end{matrix} 
\bigg]
\end{align}
where $\Gamma(.)$ is the gamma function and $H^{m,n}_{p,q}[.]$ is the univariate Fox's $H$-function defined in [11, eq. (1.2)].
\label{Proposition_1}
\begin{proposition} Let all RVs, $\gamma_i$ $\forall \in \{i,\cdots,L \}$, follow i.n.i.d. Fisher-Snedecor $\mathcal{F}$ distribution. Thus, the PDF of $\gamma=\text{max}\{\gamma_1,\cdots,\gamma_L\}$ is given as
\label{eq_3}
\begin{align}
&f_{\gamma}(\gamma)=\bigg(\prod_{i=1}^{L}\frac{\Xi_i^{m_i}}{\Gamma(m_i) \Gamma(m_{s_i})}\bigg)\nonumber\\
&\times \gamma^{\Omega-1}H^{0,1:[1,2]_{i=1:L}}_{1,1:[2,2]_{i=1:L}}
\bigg[ 
\Xi_1 \gamma,\cdots,\Xi_L \gamma\bigg\vert
\begin{matrix}
  (-\Omega;\{1\}_{i=1:L})\\
  (1-\Omega;\{1\}_{i=1:L})\\
\end{matrix}\bigg\vert \nonumber\\
&\begin{matrix}
  [(1-m_i-m_{s_i},1), (1-m_i,1)]_{i=1:L}\\
  [(0,1),(-m_i,1)]_{i=1:L}\\
\end{matrix} 
\bigg]
\end{align}
 where $\Omega=\sum_{i=1}^L m_i$ and $H^{m,n:m_1,n_1;\cdots;m_L,n_L}_{p,q:p_1,q_1;\cdots;p_L,q_L}[.]$ is the multivariate Fox's $H$-function [11, eq. (A.1)]. An efficient MATLAB code that is readily implemented by [12] to compute the multivariate Fox's $H$-function is used in this work. This because this function is not yet available as a built-in in the popular software packages such as MATLAB and MATHEMATICA.
 \end{proposition} 

\label{Proposition_1}
\begin{IEEEproof}
The CDF of the maximum i.n.i.d. variates can be computed by [1]
\label{eqn_4}
\begin{align}
F_{\gamma}(\gamma)=\prod_{i=1}^L F_{\gamma_i}(\gamma)
\end{align}
\par Substituting (2) into (4), yielding
\label{eqn_5}
\begin{align}
F_{\gamma}(\gamma)&=\prod_{i=1}^L \frac{\Xi_i^{m_i} \gamma^{m_i}}{\Gamma(m_i) \Gamma(m_{s_i})}
 \nonumber\\
&\times H^{1,2}_{2,2} 
\bigg[ \Xi_i \gamma \bigg\vert
\begin{matrix}
  (1-m_i-m_{s_i},1), (1-m_i,1)\\
  (0,1),(-m_i,1)\\
\end{matrix} 
\bigg]
\end{align}
\par After using the definition of the single variable Fox's $H$-function [11, eq. (1.2)], (5) can be expressed in multiple Barnes-type closed contours as 
 \label{eqn_6}
\begin{align}
&F_{\gamma}(\gamma)=\Bigg(\prod_{i=1}^L \frac{\Xi_i^{m_i}}{\Gamma(m_i) \Gamma(m_{s_i})}\Bigg) \frac{1}{(2 \pi j)^L} \int_{\mathbb{U}_1} \cdots \int_{\mathbb{U}_L}  \nonumber\\
& \bigg\{\prod_{i=1}^L \frac{\Gamma(u_i) \Gamma(m_i+m_{s_i}-u_i) \Gamma(m_i-u_i)}{\Gamma(1+m_i-u_i)}\bigg\} \Xi^{-u_1}_1 \cdots \Xi^{-u_L}_L \nonumber\\
& 
\gamma^{\sum_{i=1}^{L} m_i-u_i} du_1 \cdots du_L
\end{align}
where $j=\sqrt{-1}$ and $\mathbb{U}_i$ is the $i$th suitable contours in the $u$-plane from $\sigma_i-j\infty$ to $\sigma_i+j\infty$ with $\sigma_i$ is a constant value. 
 \par Differentiating (6) with respect to $\gamma$ to obtain $f_\gamma(\gamma)$, i.e. $f_\gamma(\gamma)=dF_\gamma(\gamma)/d\gamma$ and then employing the identity $\Gamma(1+x)=x\Gamma(x)$ [10, eq. (8.331.1)]. Thus, this yields  
 \label{eqn_7} 
 \begin{align}
&f_{\gamma}(\gamma)=\Bigg(\prod_{i=1}^L \frac{\Xi_i^{m_i}}{\Gamma(m_i) \Gamma(m_{s_i})}\Bigg) \frac{1}{(2 \pi j)^L} \int_{\mathbb{U}_1} \cdots \int_{\mathbb{U}_L}  \nonumber\\
& \bigg\{\prod_{i=1}^L \frac{\Gamma(u_i) \Gamma(m_i+m_{s_i}-u_i) \Gamma(m_i-u_i)}{\Gamma(1+m_i-u_i)}\bigg\} \Xi^{-u_1}_1 \cdots \Xi^{-u_L}_L\nonumber\\
& 
\frac{\Gamma(1+\sum_{i=1}^{L} m_i-u_i)}{\Gamma(\sum_{i=1}^{L} m_i-u_i)} \gamma^{\sum_{i=1}^{L} m_i-u_i-1} du_1 \cdots du_L
\end{align}
\par With the help of [11, eq. (A.1)], (7) can be written in exact closed-form expression as in (3), which completes the proof.
\end{IEEEproof} 

\label{Proposition_2}
\begin{proposition}
The MGF of $\gamma=\text{max}\{\gamma_1,\cdots,\gamma_L\}$, $\mathcal{M}_\gamma(s)$, is given as 
\label{eq_8}
\begin{align}
&\mathcal{M}_\gamma(s)=\bigg(\prod_{i=1}^{L}\frac{\Xi_i^{m_i}}{\Gamma(m_i) \Gamma(m_{s_i})}\bigg)\frac{1}{s^\Omega}\nonumber\\
&\times H^{0,1:[1,2]_{i=1:L}}_{1,0:[2,2]_{i=1:L}}
\bigg[ 
\frac{\Xi_1}{s},\cdots,\frac{\Xi_L}{s} \gamma\bigg\vert
\begin{matrix}
  (-\Omega;\{1\}_{i=1:L})\\
  -\\
\end{matrix}\bigg\vert \nonumber\\
&\begin{matrix}
  [1-m_i-m_{s_i}, 1-m_i]_{i=1:L}\\
  [0,-m_i]_{i=1:L}\\
\end{matrix} \bigg]
\end{align}
\end{proposition}

\begin{IEEEproof}
The MGF can be calculated by plugging (6) into $\mathcal{M}_\gamma(s)=s \mathcal{L} \{F_\gamma(\gamma);-s\}$ where $\mathcal{L}\{.\}$ denotes the Laplace transform. Hence, we have 
 \label{eqn_9}
\begin{align}
&\mathcal{M}_\gamma(s)=\Bigg(\prod_{i=1}^L \frac{\Xi_i^{m_i}}{\Gamma(m_i) \Gamma(m_{s_i})}\Bigg) \frac{1}{(2 \pi j)^L} \int_{\mathbb{U}_1} \cdots \int_{\mathbb{U}_L}  \nonumber\\
& \bigg\{\prod_{i=1}^L \frac{\Gamma(u_i) \Gamma(m_i+m_{s_i}-u_i) \Gamma(m_i-u_i)}{\Gamma(1+m_i-u_i)}\bigg\} \Xi^{-u_1}_1 \cdots \Xi^{-u_L}_L \nonumber\\
& 
s\mathcal{L}\{\gamma^{\sum_{i=1}^{L} m_i-u_i};-s\} du_1 \cdots du_L
\end{align}
\par The Laplace transform in (8) is recoded in [10, eq. (3.381.4)]; thus, $\mathcal{M}_\gamma(s)$ can be derived as
\label{eqn_10}
\begin{align}
&\mathcal{M}_\gamma(s)=\Bigg(\prod_{i=1}^L \frac{\Xi_i^{m_i}}{\Gamma(m_i) \Gamma(m_{s_i})}\Bigg) \frac{1}{(2 \pi j)^L} \int_{\mathbb{U}_1} \cdots \int_{\mathbb{U}_L}  \nonumber\\
& \bigg\{\prod_{i=1}^L \frac{\Gamma(u_i) \Gamma(m_i+m_{s_i}-u_i) \Gamma(m_i-u_i)}{\Gamma(1+m_i-u_i)}\bigg\} \Xi^{-u_1}_1 \cdots \Xi^{-u_L}_L \nonumber\\
& 
\frac{\Gamma(1+\sum_{i=1}^{L} m_i-u_i)}{s^{\sum_{i=1}^{L} m_i-u_i}} du_1 \cdots du_L
\end{align}  
\par Again, with the aid of [11, eq. (A.1)], (8) is deduced and the proof is accomplished.
\end{IEEEproof}

\section{Performance of SC over Non-Identically Distributed Fisher-Snedecor $\mathcal{F}$ Fading Channels}
Due to the space limitations, the following unified framework can be utilised
\label{eqn_11}
\begin{align}
&\mathcal{P}=\int_0^\infty \mathcal{P}(\gamma) f_\gamma(\gamma) d\gamma
\end{align}  
where $\mathcal{P}$ and $\mathcal{P}(\gamma)$ are the average and the conditional of the performance metric, respectively.
\par Substituting (7) into (11), we have 
 \label{eqn_12} 
 \begin{align}
&\mathcal{P}=\Bigg(\prod_{i=1}^L \frac{\Xi_i^{m_i}}{\Gamma(m_i) \Gamma(m_{s_i})}\Bigg) \nonumber\\
& \frac{1}{(2 \pi j)^L} \int_{\mathbb{U}_1} \cdots \int_{\mathbb{U}_L}   \frac{\Gamma(1+\Omega-\sum_{i=1}^{L}u_i)}{\Gamma(\Omega-\sum_{i=1}^{L}u_i)} \nonumber\\
& \bigg\{\prod_{i=1}^L \frac{\Gamma(u_i) \Gamma(m_i+m_{s_i}-u_i) \Gamma(m_i-u_i)}{\Gamma(1+m_i-u_i)}\bigg\}  \Xi^{-u_1}_1 \cdots \Xi^{-u_L}_L \nonumber\\
&  \underbrace{\int_0^\infty \gamma^{\Omega-\sum_{i=1}^{L}u_i-1} \mathcal{P}(\gamma) d\gamma}_{\mathcal{I}} du_1 \cdots du_L
\end{align}
\subsection{Average Bit Error Probability}
The ABEP can be evaluated by [1]
\label{eqn_13}
\begin{equation}
P_e=\int_0^\infty Q(\sqrt{2 \rho \gamma}) f_\gamma(\gamma) d\gamma
\end{equation} 
where $Q(.)$ is the Gaussian $Q$-function presented in [1, eq. (4.1)] and $\rho$ represents the modulation parameter. For example, $\rho = 1$ for binary phase shift keying (BPSK), while $\rho = 0.5$ for binary frequency shift keying (BFSK).
\par Inserting (7) in (13) and invoking the identity [13, eq. (13)], $\mathcal{I}$ of (12) is obtained as 
\label{eqn_14}
\begin{align}
\mathcal{I}&=\frac{1}{2 \sqrt{\pi}}\int_0^\infty \gamma^{\Omega-\sum_{i=1}^{L} u_i-1} H^{2,0}_{1,2}
\bigg[ 
\rho \gamma \bigg\vert
\begin{matrix}
  (1,1)\\
  (0,1),(0.5,1)\\
\end{matrix} \bigg] d\gamma \nonumber\\
&\stackrel{(a)}{=} \rho^{-\Omega+\sum_{i=1}^{L}u_i} \frac{\Gamma(\Omega-\sum_{i=1}^{L}u_i) \Gamma(0.5+\Omega-\sum_{i=1}^{L}u_i)}{\Gamma(1+\Omega-\sum_{i=1}^{L}u_i)}
\end{align}
where $(a)$ follows [11, eq. (2.8)]. 
\par Next, plugging (14) in (12), performing some mathematical straightforward simplifications and using [11, eq. (A.1)], $P_e$ is obtained as  
\label{eq_15}
\begin{align}
&P_e=\bigg(\prod_{i=1}^{L}\frac{\Xi_i^{m_i}}{\Gamma(m_i) \Gamma(m_{s_i})}\bigg)\frac{1}{2\sqrt{\pi} \rho^\Omega}\nonumber\\
&\times H^{0,1:[1,2]_{i=1:L}}_{1,0:[2,2]_{i=1:L}}
\bigg[ 
\frac{\Xi_1}{\rho},\cdots,\frac{\Xi_L}{\rho}\bigg\vert
\begin{matrix}
  (0.5-\Omega;\{1\}_{i=1:L})\\
 -\\
\end{matrix}\bigg\vert \nonumber\\
&\begin{matrix}
  [(1-m_i-m_{s_i},1), (1-m_i,1)]_{i=1:L}\\
  [(0,1),(-m_i,1)]_{i=1:L}\\
\end{matrix} \bigg]
\end{align}
\subsection{Average Channel Capacity}
According to Shannon theory, the ACC, $\bar{C}$, can be computed by
\label{eqn_16}
\begin{equation}
\bar{C}=\frac{B}{\text{ln}2}\int_0^\infty \text{ln}(1+\gamma) f_\gamma(\gamma) d\gamma
\end{equation} 
where $B$ is the bandwidth of the channel.
\par Inserting (7) in (16), $\mathcal{I}$ of (12) for $\bar{C}$ becomes
\label{eqn_17}
\begin{align}
\mathcal{I}&=\frac{B}{\text{ln}2} \int_0^\infty \gamma^{\Omega-\sum_{i=1}^L u_i-1} \text{ln}(1+\gamma)d\gamma \nonumber\\
&\stackrel{(b)}{=} \frac{B}{\text{ln}2} \frac{ \Gamma(1-\Omega+\sum_{i=1}^L u_i) [\Gamma(\Omega-\sum_{i=1}^L u_i)]^2}{\Gamma(1+\Omega-\sum_{i=1}^L u_i)}
\end{align}
where $(b)$ follows after employing [10, eq. (4.293.10)] and making use of the properties [10, eq. (8.334.3)] and [10, eq. (8.331.1)].
 \par Now, substituting $(b)$ of (17) into (12) and doing some algebraic manipulations, $\bar{C}$ is yielded as follows
\label{eq_18}
\begin{align}
&\bar{C}=\bigg(\prod_{i=1}^{L}\frac{\Xi_i^{m_i}}{\Gamma(m_i) \Gamma(m_{s_i})}\bigg)\frac{B}{\text{ln}2} H^{1,1:[1,2]_{i=1:L}}_{1,1:[2,2]_{i=1:L}}
\bigg[\Xi_1,\cdots,\Xi_L\bigg\vert \nonumber\\
&
\begin{matrix}
  (1-\Omega;\{1\}_{i=1:L})\\
 (1-\Omega;\{1\}_{i=1:L})\\
\end{matrix}\bigg\vert 
\begin{matrix}
  [(1-m_i-m_{s_i},1), (1-m_i,1)]_{i=1:L}\\
  [(0,1),(-m_i,1)]_{i=1:L}\\
\end{matrix} \bigg]
\end{align}
\par It can be noted that (18) reduces to [15, eq. (18)] for $L=1$.
\subsection{ED with SC over Fisher-Snedecor $\mathcal{F}$ fading conditions}
\subsubsection{Average Detection Probability}
\par The ADP can be evaluated by [9, eq. (9)/eq. (4)]
\label{eqn_19}
\begin{align}
\bar{P}_d=\int_0^\infty Q_u(\sqrt{2 \gamma}, \sqrt{\lambda}) f_\gamma(\gamma) d\gamma
\end{align}
where $\lambda$ is the threshold value, $u=TW$ stands for the time-bandwidth product and $Q_u(., .)$ is the generalized Marcum $Q$-function.
\label{eq_22}
\setcounter{equation}{21}
\begin{table*}[h]
\begin{align}
\bar{P}_d=&1-\pi\bigg(\frac{\lambda}{2}\bigg)^u \bigg(\prod_{i=1}^{L}\frac{\Xi_i^{m_i}}{\Gamma(m_i) \Gamma(m_{s_i})}\bigg) H^{0,3:1,0;1,0;[1,2]_{i=1:L}}_{3,2:0,1;1,3;[2,2]_{i=1:L}}
\bigg[ \frac{\lambda}{2},\frac{\lambda}{2},\Xi_1,\cdots,\Xi_L \bigg\vert\begin{matrix}
  (1-u;1,1,\{0\}_{i=1:L}), (1-\Omega;0,1,\{1\}_{i=1:L})\\
 (-u;1,1,\{0\}_{i=1:L})\\
\end{matrix}\bigg\vert \nonumber\\
&\begin{matrix}
  (-\Omega;0,0,\{1\}_{i=1:L})\\
 (1-\Omega;0,0,\{1\}_{i=1:L})\\
\end{matrix}\bigg\vert 
\begin{matrix}
  -\\
  (0,1)\\
\end{matrix} \bigg\vert
\begin{matrix}
  (0.5,1)\\
  (0,1),(1-u,1),(0.5,1)\\
\end{matrix} \bigg\vert
\begin{matrix}
  [(1-m_i-m_{s_i},1), (1-m_i,1)]_{i=1:L}\\
  [(0,1),(-m_i,1)]_{i=1:L}\\
\end{matrix} \bigg]
\end{align}
\hrulefill
\vspace*{1pt}
\label{eq_26}
\setcounter{equation}{25}
\begin{align}
&\bar{A}=1-\bigg(\prod_{i=1}^{L}\frac{\Xi_i^{m_i}}{\Gamma(m_i) \Gamma(m_{s_i})}\bigg)\sum_{k=0}^{u-1}\sum_{l=0}^{k}{{k+u-1}\choose{k-l}}\frac{1}{2^{k+\Omega+u}l!}H^{0,2:[1,2]_{i=1:L}}_{2,1:[2,2]_{i=1:L}}
\bigg[ 2\Xi_1,\cdots,2\Xi_L \bigg\vert \nonumber\\
& \hspace{4 cm}
\begin{matrix}
  (-\Omega,\{1\}_{i=1:L}),(1-l-\Omega,\{1\}_{i=1:L})\\
 (1-\Omega,\{1\}_{i=1:L})\\
\end{matrix} \bigg\vert 
\begin{matrix}
  [(1-m_i-m_{s_i},1), (1-m_i,1)]_{i=1:L}\\
  [(0,1),(-m_1,1)]_{i=1:L}\\
\end{matrix} \bigg]
\end{align}
\hrulefill
\vspace*{1pt}
\end{table*}
\par It can be observed that the generalized Marcum $Q$-function can be expressed as
\label{eqn_20}
\setcounter{equation}{19}
\begin{align}
&Q_u(\sqrt{2 \gamma}, \sqrt{\lambda})\nonumber\\
&  \stackrel{(c_1)}{=} 1-\frac{e^{-\gamma}}{2^{\frac{u+1}{2}} \gamma^{\frac{u-1}{2}}}  \int_0^\lambda x^{\frac{u-1}{2}} e^{-\frac{x}{2}} I_{u-1}(\sqrt{2\gamma x}) dx  \nonumber\\
&\stackrel{(c_2)}{=} 1- \frac{ \pi e^{-\gamma}}{2^u } \int_0^\lambda x^{u-1} 
H^{1,0}_{0,1}
\bigg[ 
 \frac{x}{2} \bigg\vert
\begin{matrix}
 -\\
  (0,1)\\
\end{matrix}  \bigg] \nonumber\\
&
\times H^{1,0}_{1,3}
\bigg[ 
 \frac{\gamma x}{2} \bigg\vert
\begin{matrix}
 (0.5,1)\\
  (0,1),(1-u,1),(0.5,1)\\
\end{matrix}\bigg] dx
\nonumber\\
&\stackrel{(c_3)}{=}1-\pi\bigg(\frac{\lambda}{2}\bigg)^u e^{-\gamma} \frac{1}{(2 \pi j)^2}\int_{\mathbb{R}_1} \int_{\mathbb{R}_2} \frac{\Gamma(u-r_1-r_2)}{\Gamma(1+u-r_1-r_2)}\nonumber\\
&\frac{\Gamma(r_1) \Gamma(r_2)}{\Gamma(0.5+r_2) \Gamma(u-r_2) \Gamma(0.5-r_2)}   \bigg(\frac{\lambda}{2}\bigg)^{-r_1} \bigg(\frac{\lambda \gamma}{2}\bigg)^{-r_2} dr_1 dr_2
\end{align}
where $(c_1)$ and $(c_2)$ arise after employing [1, eq. (4.60)] and then respectively utilising the properties [11, eq. (1.39)] and [14, eq. (03.02.26.0067.01)] for the exponential function and $I_{a}(.)$, which represents the modified Bessel function of the first kind and $a$th-order. Using the definition of the univariate Fox's $H$-function [11, eq. (1.2)] and solving the integral of $(c_2)$, then $(c_3)$ follows in terms of the contour integral form where $\mathbb{R}_1$ and $\mathbb{R}_2$ are the suitable closed contours in the complex $r$-plane.
\par Now, plugging (7) and (20) into (19) and using the fact that $\int_0^\infty f_\gamma(\gamma)d\gamma \triangleq1$, $\mathcal{I}$ of (12) is deduced as follows
\label{eqn_21}
\begin{align}
&\mathcal{I}=1-\pi\bigg(\frac{\lambda}{2}\bigg)^u e^{-\gamma} \frac{1}{(2 \pi j)^2}\int_{\mathbb{R}_1} \int_{\mathbb{R}_2} \frac{\Gamma(u-r_1-r_2)}{\Gamma(1+u-r_1-r_2)}\nonumber\\
&\frac{\Gamma(r_1) \Gamma(r_2)}{\Gamma(0.5+r_2) \Gamma(u-r_2) \Gamma(0.5-r_2)}   \bigg(\frac{\lambda}{2}\bigg)^{-r_1} \bigg(\frac{\lambda \gamma}{2}\bigg)^{-r_2} \nonumber\\
& \int_0^\infty \gamma^{\Omega-\sum_{i=1}^{L}u_i-r_2-1} e^{-\gamma} dr_1 dr_2
\end{align}
\par Recalling [10, eq. (3.381.4)] for the inner integral of (21), substituting the result into (12) and making employ of [11, eq. (A.1)], then $\bar{P}_d$ is obtained as shown on the top of this page.
\par In contrast to [9, eq. (14)] and [16, eq. (14)] that are derived for no diversity scenario in terms of the infinite series, (22) for $L=1$ can be obtained in exact closed-from computationally tractable expression in terms of a single variable Fox's $H$-function.  
\subsubsection{Average AUC}
The average AUC is a single figure of merit that can be used in the analysis of performance of the ED when the plotting of the ADP versus the probability of false alarm, namely, ROC, doesn't provide a clear insight into the behaviour of the system. 
\par The average AUC, $\bar{A}$, can be calculated by [9, eq. (36)]
 \label{eqn_23}
 \setcounter{equation}{22}
\begin{align}
&\bar{A}=\int_0^\infty A(\gamma) f_\gamma(\gamma)d\gamma
\end{align}
where $A(\gamma)$ is the AUC at the instantaneous SNR.
\par The $A(\gamma)$ is given as [9, eq. (35)]
 \label{eqn_24}
\begin{align}
&A(\gamma)=1-\sum_{k=0}^{u-1}\sum_{l=0}^{k}{{k+u-1}\choose{k-l}}\frac{1}{2^{k+l+u}l!}\gamma^l e^{-\frac{l}{2}}
\end{align}
where ${{b}\choose{a}}$ denotes the binomial coefficient.
\par Substituting (24) and (7) into (23) and invoking $\int_0^\infty f_\gamma(\gamma)d\gamma \triangleq1$, we have $\mathcal{I}$ of (12) as 
\label{eqn_25}
\begin{align}
&\mathcal{I}=1-\sum_{k=0}^{u-1}\sum_{l=0}^{k}{{k+u-1}\choose{k-l}}\frac{1}{2^{k+l+u}l!} \nonumber\\
& \times \int_0^\infty \gamma^{l+\Omega-\sum_{i=1}^{L}u_i-r_2-1} e^{-\frac{l}{2}} d\gamma
\end{align}
\par Utilising [10, eq. (3.381.4)] to evaluate the integral of (25) and plugging the result in (12), we have a closed-form expression of $\bar{A}$ as given on the top of this page.

\begin{figure}[t]
\centering
  \includegraphics[width=3.5 in, height=2.5 in]{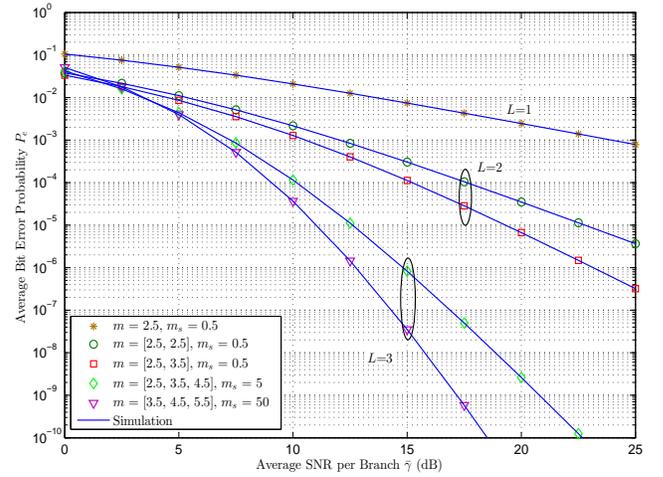}
\centering
\caption{ABEP for BPSK comparison between single receiver, dual and triple i.n.i.d. branches of SC versus $\bar{\gamma}$ for different $m$ and $m_{s}$.}
\end{figure}

\section{Analytical and Simulation Results}
In this section, to validate our derived PDF and MGF of the maximum of i.n.i.d. Fisher-Snedecor $\mathcal{F}$ variates, the ABEP, the ACC, the ADP, and the average AUC of SC diversity are analysed. The Monte Carlo simulations that are obtained via generating $10^7$ realizations for each RV are compared with the analytical results. In all figures, the multivariate Fox's $H$-function has been evaluated by the MATLAB code that was implemented by [12]. Additionally, the solid lines corresponds to the simulations results whereas the markers represents the numerical results. Three scenarios of the shadowing impact, which are heavy, moderate, and light shadowing are studied by using $m_s = 0.5$, $m_s = 5$ and $m_s = 50$, respectively.
\begin{figure}[t]
\centering
  \includegraphics[width=3.5 in, height=2.5 in]{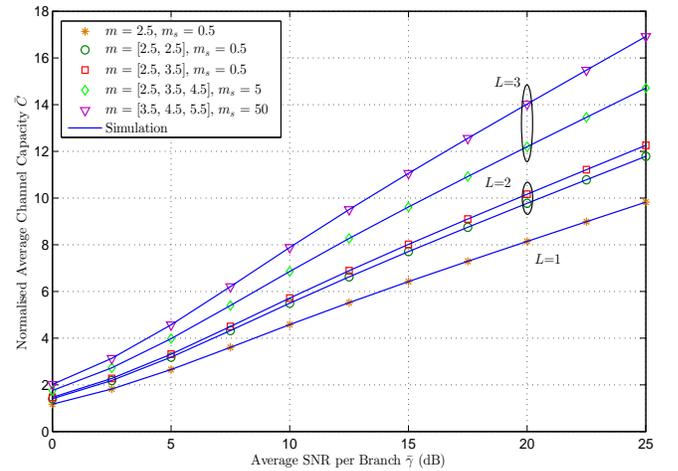}
\centering
\caption{Normalised ACC comparison between single receiver, dual and triple i.n.i.d. branches of SC versus $\bar{\gamma}$ for different $m$ and $m_{s}$.}
\end{figure}

\begin{figure}[t]
\centering
  \includegraphics[width=3.5 in, height=2.5 in]{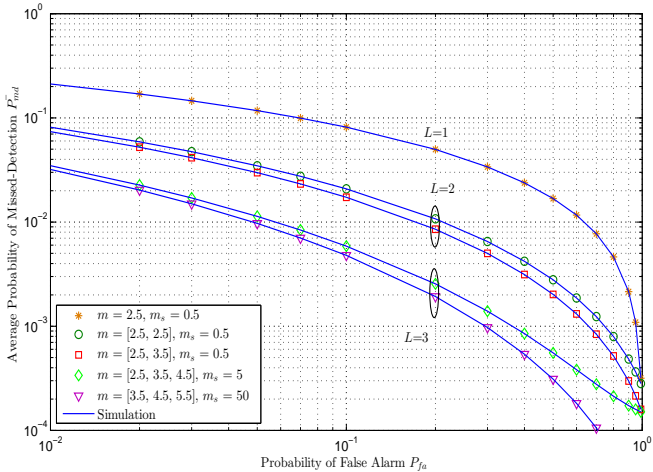}
\centering
\caption{Complementary ROC comparison between single receiver, dual and triple i.n.i.d. branches of SC for $u = 3$, $\bar{\gamma} = 15$ dB and different $m$ and $m_{s}$.}
\end{figure} 

\begin{figure}[t]
\centering
  \includegraphics[width=3.5 in, height=2.5 in]{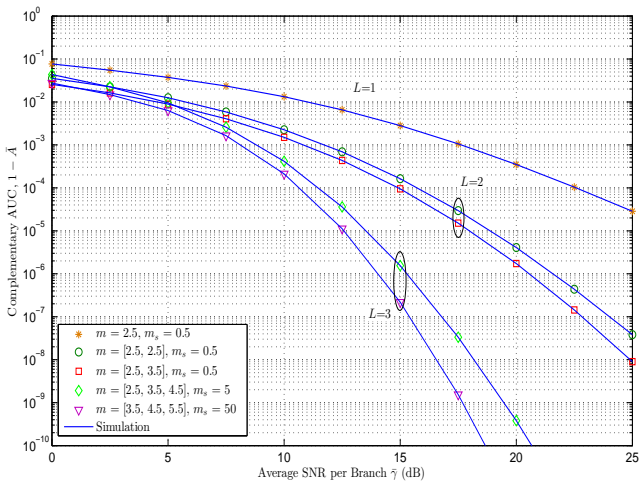} 
\centering
\caption{Complementary AUC comparison between single receiver, dual and triple i.n.i.d. branches of SC versus $\bar{\gamma}$ for $u = 3$ and different $m$ and $m_{s}$.}
\end{figure}   
\par Figs. 1, 2, and 4 illustrate the ABEP for BPSK, the normalised ACC, and the complementary AUC ($1-\bar{A}$) with single receiver, dual, and triple SC branches over i.n.i.d. Fisher-Snedecor $\mathcal{F}$ fading channels versus the average SNR per branch, $\bar{\gamma}$, respectively, for different scenarios of the fading parameters. In the same context, Fig. 3 explains the complementary ROC, which plots the average probability of missed-detection ($1-\bar{P}_d$) versus the probability of false alarm $P_f(\lambda)=\Gamma(u,\lambda/2)/\Gamma(u)$ for $u=3$ and $\bar{\gamma} = 15$ dB\footnote{Here, $\Gamma(.,.)$ represents the upper incomplete gamma function [10, eq. (8.350.2)].}. As anticipated, the performance of the communication systems becomes better when the SC diversity is employed and monotonically improves with the increasing in the number of diversity branches. The reason has been widely presented in the literature, which is the received average SNR of SC scheme is higher than the no-diversity and its increases when $L=3$ is used rather than $L=2$. For comparison purpose, the scenario $m=[3.5, 4.5, 5.5]$ and $m_s =50$ that was studied in [7, Fig. 3], has been utilised here. As expected, the MRC diversity provides less ABEP than the SC branches but with high implementation complexity.  
\par In all provided figures, the perfect matching between the numerical results and their Monte Carlo simulation counterparts can be observed, which confirms the validation of our derived expressions.

\section{Conclusions}
In this paper, the PDF and the MGF of the maximum of not necessarily identically distributed Fisher-Snedecor $\mathcal{F}$ RVs were derived in terms of the multivariate Fox's $H$-function that has been widely used and implemented in the literature. These statistics were then employed to analyse the performance of SC diversity with non-identically distributed branches. To be specific, the ABEP, the ACC, the ADP, and the AUC of ED technique were obtained in exact mathematically tractable closed-form expressions. Comparisons of our results with previous works that were achieved by using a single receiver and MRC scheme as well as the numerical and simulation results for different scenarios have been carried out via using the same simulation parameters.

\ifCLASSOPTIONcaptionsoff
  \newpage
\fi

\end{document}